\documentclass[prd,aps,twocolumn,showpacs,tightenlines]{revtex4}
\usepackage{amsmath}
\usepackage{color,graphicx}
\usepackage{epsfig}

\def\Journal#1#2#3#4{{#1} {\bf #2}, #3 (#4)}

\def\NPA{Nucl. Phys. A}
\def\NPB{Nucl. Phys. B}

\def\PLB{Phys. Lett. B}
\def\PRL{Phys. Rev. Lett.}

\def\PRD{Phys. Rev. D}
\def\ZPC{Z. Phys. C}
\def\EPJ{Eur. Phys. Journal}

\begin{document}

\title{Parton-Hadron Duality in Unpolarised and Polarised Structure Functions}

\author{N. Bianchi}
\email[]{bianchi@lnf.infn.it}
\affiliation{Laboratori Nazionali di Frascati dell'INFN, Via E. Fermi 40, 
00044 Frascati (RM), Italy}

\author{A. Fantoni}
\email[]{fantoni@lnf.infn.it}
\affiliation{Laboratori Nazionali di Frascati dell'INFN, Via E. Fermi 40, 
00044 Frascati (RM), Italy}

\author{S.~Liuti}
\email[]{sl4y@virginia.edu}
\affiliation{University of Virginia, Charlottesville, Virginia 22901, USA.}

\date{
\today}

\begin{abstract}
We study the phenomenon of parton-hadron duality in both polarised 
and unpolarised electron proton scattering using the HERMES and the 
Jefferson Lab data, respectively. 
In both cases we extend a systematic perturbative QCD based analysis
to the integrals of the structure functions in the resonance region.  
After subtracting target mass corrections 
and large $x$ resummation effects,  
we extract the remaining power corrections up to order $1/Q^2$. 
We find a sizeable suppression of these terms with respect to analyses
using deep inelastic scattering data. The suppression appears consistently
in both polarised and unpolarised data, except for the low $Q^2$ polarised
data, where a large negative higher twist contribution remains.     
Possible scenarios generating this behavior are discussed.       
\end{abstract}

\maketitle

\section{Introduction}

The structure of hadrons and their interactions can be described within two
different but complementary approaches
based on either partonic or hadronic degrees 
of freedom. The first one is expected to be valid at high energy, 
while the second one is applicable at low energy     
where the effects of confinement become large.
In some specific cases where a description in terms of non-partonic degrees 
of freedom seems more natural, the quark-gluon description can be also 
successfully used.  
This observation is called parton-hadron duality. 
It was introduced for deep inelastic scattering (DIS) 
by Bloom and Gilman \cite{BG} who reported 
an equivalence between the smooth $x$  dependence of the inclusive structure 
function at large 
$Q^2$ and the average over $W^2$ of the nucleon resonances  
($x=Q^2/2M\nu$, $Q^2$ 
is the four momentum transfer 
squared, $M$ is the nucleon mass, $\nu$ is the energy transfer, and 
$W^2 = Q^2(1/x-1)+M^2$ is the final state invariant mass). 
One refers to {\em global} duality if the average, defined {\it e.g.} as 
the integral of the structure functions, 
is taken over the over the whole resonance region 
$ 1 \leq W^2 \leq 4$ GeV$^2$. If, however, 
the averaging is performed over smaller 
$W^2$ ranges, extending {\it e.g.} over single resonances, 
one can analyze the onset of {\em local} duality. 


More generally, the concept of duality is often assumed in QCD-based 
interpretations 
of most hard scattering experiments, such as DIS, 
$e^+e^-$ annihilation into hadrons, and hadron-hadron collisions.   
Its usage appears whenever hadronic 
observables (mostly averaged over a given energy range) 
are replaced by calculable partonic ones with
little more going into the hadronic formation phase 
of each process  -- from partons to hadrons or vice versa.
In a phenomenological context, duality studies are aimed  
at establishing to what extent a partonic description of the hard 
scattering process can determine the structure of the final state. 
In fact, as prescribed by the factorization property of QCD, we visualize 
hard scattering processes happening in two stages, one dominated by short 
times and distances and involving only parton jets, followed by hadron
formation at a much larger scale. Duality is intrinsic to the factorization
property. Violations of duality might signal violations of factorization
in that, for instance, the probe-parton interaction might occur at larger
time scales than required in order to exclude parton (re)interactions.   

With the advent of both more detailed studies of soft scales and 
confinement \cite{reticolo}, and higher precision measurements covering
a wide range of reactions, it is now becoming 
possible to investigate the role of duality in QCD as a 
subject per se. 
For example, recent studies 
of local parton-hadron duality and 
its violations in semi-leptonic decays, and $\tau$ decays 
illustrate how the possible impact of these experiments on the 
extraction 
of the Cabibbo-Kobayashi-Maskawa (CKM) matrix elements, depends on the  
size of violations of local 
duality \cite{CKM}. A practical necessity to address duality 
quantitatively exists also for inclusive $ep$ scattering where 
most of the currently available large $x$ data lie in the resonance 
region. In fact, for  
$x >0.5$ and $Q^2 \gtrsim$ 5 GeV$^2$ -- a typical starting
value for perturbative evolution --  
$W^2 \leq 5$ GeV$^2$. 
Therefore, the behavior of the nucleon structure functions in the 
resonance region needs to be addressed in detail  
in order to be able to discuss 
theoretical predictions in the limit $x \rightarrow 1$. 

The first QCD-based studies of Bloom and Gilman duality 
reinterpreted the ``averaging'' procedure in terms of Mellin moments 
of the structure function.
The moments taken in the low $Q^2$ and in the DIS regime, respectively,
were shown to be equivalent to one another within the given
range and precision of the data, modulo 
perturbative corrections and relatively small power corrections \cite{DGP}.
It was conjectured that duality resulted from a 
cancellation of higher order 
terms in the twist expansion 
that would otherwise be expected to 
dominate the cross section at $x \rightarrow 1$.
This view has been adopted since, particularly in the more recent 
studies in Ref.~\cite{twistpapers}. 
In Ref.~\cite{SIMO1} a new analysis was performed, using 
the recent inclusive unpolarised electron-nucleon scattering 
data on hydrogen and deuterium targets from Jefferson Lab \cite{CEBAF}. 
It was shown in particular that, because of the increased precision
of the data, one is now able to unravel different sources of 
scaling violations affecting the structure functions, 
namely Target Mass Corrections
(TMC), Large $x$ Resummation effects (LxR), and dynamical Higher Twists (HTs), 
in addition to the standard Next-to-Leading-Order (NLO) perturbative
evolution. As a result, contrarily to what originally deduced in 
{\it e.g.} Ref.\cite{CEBAF}, a more pronounced role of the HT terms 
is obtained, pointing at the fact that duality, defined on the basis of a
dominance of single parton scattering, could indeed be broken. 
 
In contrast to the extensive study of duality for the unpolarised,
{\it i.e.} spin averaged, photo-absorption cross section, 
the validity of duality 
has not been investigated until very recently 
for the spin structure function $g_1$, which is 
proportional to the spin-{\it dependent} photo-absorption cross section. 
Evidence of duality for the spin asymmetry $A_1$ was reported in 
Ref.~\cite{HERMESPRL}. 
A phenomenological study addressing parton-duality 
was performed in \cite{EPKW} using the low $Q^2$ 
data from Ref.~\cite{E143}.   
Studies of duality for $g_1$ are of particular interest 
because they might help understanding the transition from 
the large $Q^2$ regime described by pQCD, and the $Q^2 \rightarrow 0$
limit, where the Gerasimov-Drell-Hearn sum rule is 
expected to apply \cite{GDH}. 
They may also lead
to a complementary method to study the spin structure of the nucleon
at large $x$, which is difficult to measure in the DIS region with
high statistics. In particular, they might provide additional 
information on the transition from single parton scattering, to
the dominance of processes where several partons are involved 
\cite{BroTemple}. In this respect, it is important to perform an 
analysis aimed at disentangling the different contributions 
to the $Q^2$ dependence of $g_1$ in the resonance region.  
The aim of this paper is to carry out such an analysis by 
investigating quantitatively the onset of duality
and its violations both for the unpolarised and polarised 
structure functions.  

In Section II we define the concept of duality and we illustrate the role
of different kinematical regions; in Section III we present our analysis
and we describe our results. In particular, 
we compare the data with perturbative-QCD predictions and we
discuss in detail the contribution of different types of corrections
in both the unpolarised and polarised case. 
Finally, in Section IV we draw our conclusions.   

\section{Definitions and Kinematics}
Parton-hadron duality in DIS was first observed more than 
three decades ago. Since then it has been necessary to give new 
definitions of the quantities involved which can be described 
within QCD-based approaches. In what follows we list all such
definitions.

\subsection{Kinematical variables}
\begin{figure}[h]
   \begin{center}
      \mbox{
           \epsfysize=8cm
           \epsfbox{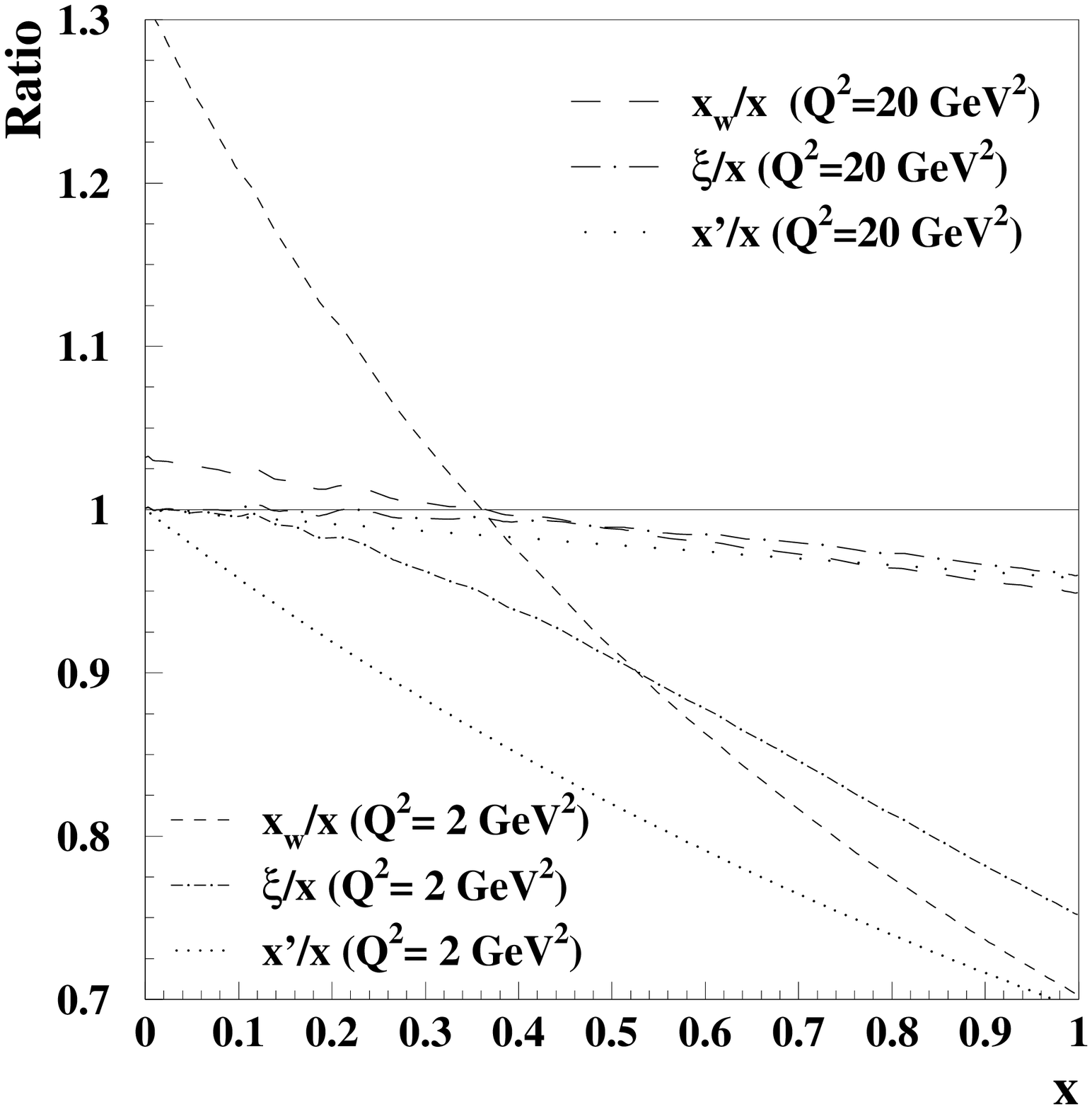}
           }
   \end{center}
\caption{Ratio between the three different variables $x'$, $\xi$ and $x_W$ 
         defined in the text and the Bjorken variable $x$ as a function of $x$.}
\label{variables}
\end{figure}

Besides the scaling variable $x$, other
variables have been used in the literature to study duality.   
A number of parameterisations based on these 
variables have been proposed that reproduce in an effective 
way some of the corrections to the perturbative QCD calculations
that we study in this paper.
The most extensively used variables are: 
$x'=1/\omega'$, where $\omega'= 1/x + M^2/Q^2$. $x'$ was originally 
introduced by Bloom and Gilman in order 
to obtain a better agreement between DIS and the resonance region; 
$\xi=2x/(1+(1+4x^2M^2/Q^2)^{1/2})$ \cite{NAC}, 
originally introduced to take into account the target mass 
effects; 
$x_w=Q^2+B/(Q^2+W^2-M^2+A)$, $A$ and $B$ being fitted parameters,  
used in Refs.~\cite{SzcUle,BOYA}. 
These additional variables include 
a $Q^2$ dependence that phenomenologically absorbs some of
the scaling violations that are important  
at low $Q^2$.
In Fig.\ref{variables}  
we compare their behavior vs. $x$ for different values of $Q^2$.
From the figure one can see that by calculating $F_2$ in $\xi$ and $x'$, 
one effectively ``rescales'' the structure function to lower values of $x$,
in a $Q^2$ dependent way, namely the rescaling is larger at lower $Q^2$.  

In this paper we present results in terms of $x$ and $Q^2$ and we 
illustrate the 
contributions of scaling violations of different origin on a case 
by case basis.

\subsection{Unpolarised structure function.}
The inclusive DIS cross section of unpolarised electrons off
an unpolarised proton is written in terms of the two structure
functions $F_2$ and $F_1$, 
\begin{eqnarray}
\label{xsect}
\frac{d^2\sigma}{dx dy} =
\frac{4\pi\alpha^2}{Q^2 xy}
\left[
    \left(1-y-\frac{(Mxy)^2}{Q^2}\right)F_2 +
    y^2 x F_1
\right], 
\end{eqnarray}
where $y=\nu/E$, $E$ being the initial electron energy.
The structure functions are related by the equation:  
\begin{equation}
\label{R}
F_1 = F_2(1+\gamma^2)/(2x(1+R)),
\end{equation}
where $\gamma^2=4M^2x^2/Q^2$, and $R$ is ratio of the longitudinal to transverse
virtual photo-absorbtion cross sections.  

In QCD, $F_2$ is expanded in series of inverse powers of $Q^2$, 
obtained by ordering the matrix 
elements in the DIS process by increasing twist $\tau$, which is equal
to their dimension minus spin:
\begin{eqnarray}
\label{t-exp} 
F_{2}(x,Q^2) = F_{2}^{LT}(x,Q^2) +
\frac{H(x,Q^2)}{Q^2} + {\cal O}\left(1/Q^4 \right)
\end{eqnarray}
The first term is the leading twist (LT), with $\tau=2$.
The terms of order $1/Q^{\tau-2}$, $\tau \geq 4$, in Eq.(\ref{t-exp}) 
are the higher order terms, generally referred to as 
higher twists (HTs). 
Additional corrections to the LT part 
due to the finite mass of the initial nucleon -- the target mass
corrections (TMC) -- are included directly in $F_2^{LT}$.
For $Q^2$ larger than $\approx $ 1 GeV$^2$, TMC are taken into
account through the following expansion \cite{GeoPol}: 
\begin{eqnarray}
\label{TMC}
F_{2}^{LT(TMC)}(x,Q^2) & = &
    \frac{x^2}{\xi^2\gamma^3}F_2^{\mathrm{Asy}}(\xi,Q^2) + \\ & &
    6\frac{x^3M^2}{Q^2\gamma^4}\int_\xi^1\frac{d \xi'}{{\xi'}^2} \nonumber
F_2^{\mathrm{Asy}}(\xi',Q^2),
\end{eqnarray}
where $F_2^{Asy}$
is the structure function in the absence of TMC. Since TMC should in 
principle be applied also to the HT, we disregard 
terms of ${\cal O}(1/Q^4)$ \cite{AKL}.  
$H$, then, represents the ``genuine'' HT correction that involves
interactions between the struck parton and the spectators or, formally,
multi-parton correlation functions.   
%

Parton-hadron duality in DIS is studied 
by considering integrals of the structure function defined as:
\begin{equation}
\label{Iexp}
I^{\mathrm{res}}(Q^2) = \int^{x_{\mathrm{max}}}_{x_{\mathrm{min}}} 
F_2^{\mathrm{res}}(x,Q^2) \; dx
\end{equation}
where $F_2^{\mathrm{res}}$ is evaluated using the experimental data  
in the resonance region. 
For each $Q^2$ value: 
$x_{\mathrm{min}}=Q^2/(Q^2+W_{\mathrm{max}}^2-M^2)$, and 
$x_{\mathrm{max}}=Q^2/(Q^2+W_{\mathrm{min}}^2-M^2)$. 
$W_{\mathrm{min}}$  and $W_{\mathrm{max}}$ delimit the resonance region. 
The same expression is then 
calculated in the same range of 
$x$ and for the same value of $Q^2$, using parameterisations 
of $F_2$ that reproduce the DIS behaviour
of the data at large $Q^2$. These
parameterisations are very well constrained in the region of interest 
($x >0.3$) although they do not correspond directly to measured data.  
Here in fact $F_2$ is dominated by the valence contribution.    
On the contrary, by using the same procedure at low $x$ where the singlet 
and gluon distributions govern $F_2$,  
one would find much larger uncertainties  
in the initial low $Q^2$ parameterisations
because of their strong correlation with the value of $\alpha_S$.
We present two forms for the DIS integrals:
\begin{equation}
I^{LT}(Q^2) = \int^{x_{\mathrm{max}}}_{x_{\mathrm{min}}} F_2^{LT}(x,Q^2) \; dx,
\label{ILT}
\end{equation}
and 
\begin{equation}
I^{HT}(Q^2) = \int^{x_{\mathrm{max}}}_{x_{\mathrm{min}}}
\bigg(F_{2}^{LT}(x,Q^2) + \frac{H(x,Q^2)}{Q^2}\bigg) \; dx.
\label{IHT}
\end{equation}

Duality is attained strictly only if the ratio:
\begin{equation}
R_{\mathrm{unpol}}^{LT}= \frac{I^{\mathrm{res}}}{I^{LT}},
\label{RdualLT}
\end{equation}
is unity. However, one can extend this definition to  
the ratio: 
\begin{equation}
R_{\mathrm{unpol}}^{HT}= \frac{I^{\mathrm{res}}}{I^{HT}} .
\label{Rduality}
\end{equation}

The rational for this
definition is that even when including the first few terms of the 
twist expansion, one is embracing a partonic description of the proton. 

\begin{figure}[h]
   \begin{center}
      \mbox{
           \epsfysize=8cm
           \epsfbox{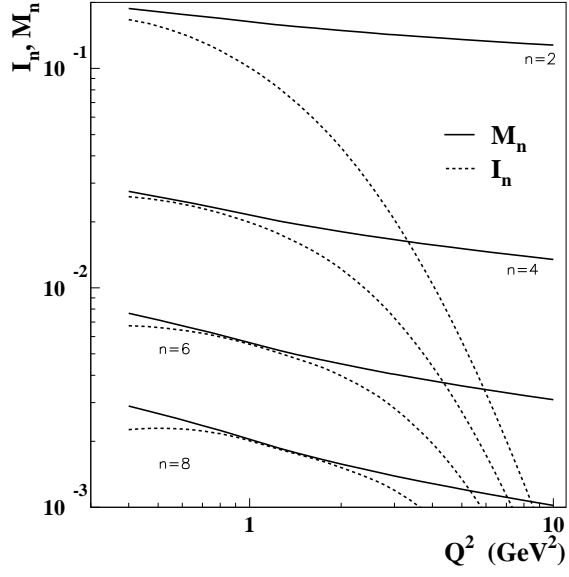}
           }
   \end{center}
\caption{The integral $I_n$ defined in the text
plotted vs. $Q^2$ (dashed line), compared to the Mellin moments 
defined in Eq.(\ref{CN}) (full lines). 
All quantities have been calculated 
for illustration using the parton distributions function 
parameterisation from Ref.~\cite{CTEQ}. 
}
\label{moments}
\end{figure}

Perturbative QCD analyses use the 
Mellin moments of the structure function,
which allow for a direct comparison with theoretical predictions. 
These are defined as:
\begin{equation}
M_n(Q^2) = {\int_0^1} dx x^{n-2} F_2(x,Q^2),
\label{CN}
\end{equation}
and by 
\begin{eqnarray}
M_n^{TMC}(Q^2) &  = & {\displaystyle \int_0^1} dx \, \xi^{n-1} \, \frac{F_2(x,Q^2)}{x} \, p_n\left(\frac{\xi}{x} \right), 
\\  \displaystyle  p_n  & = & 1 + \frac{6(n-1)}{(n+2)(n+3)} \left( \frac{\xi}{x}-1 \right)
\nonumber \\ 
& & + \frac{n(n-1)}{(n+2)(n+3)} \left( \frac{\xi}{x} -1 \right)^2, 
\label{Nach}
\end{eqnarray}
which takes into account TMC \cite{NAC}. However, one needs in this case 
experimental values of the structure function in kinematics outside the 
resonance region. This procedure renders the comparison between 
theory and experiment less straightforward. The difference
between the Mellin moments (Eqs.\ref{CN},\ref{Nach}) and the integrals
over the resonance region -- $I_n=\int^{x_{\mathrm{max}}}_{x_{\mathrm{min}}} x^{n-2} F_2^{LT}(x,Q^2)$ dx -- is shown in Fig.\ref{moments}. 
The drop of the quantities $I_n$ with respect to $M_n$ at larger values of $Q^2$, is
due to the pQCD evolution of $F_2$, that moves strength to lower values of $x$, 
outside the range $[x_{\mathrm{max}},x_{\mathrm{min}}]$.  
In our approach we use the integrals defined in Eqs.\ref{ILT},\ref{IHT}. As it can 
be understood also from the trend shown in Fig.\ref{moments}, 
these effectively describe duality also as a function of the average 
value of $x$ in each interval 
$[x_{\mathrm{min}}(Q^2), x_{\mathrm{max}}(Q^2)]$. This 
corresponds to: $\langle x \rangle = x(W^2 \equiv 2.5 \, \rm{GeV}^2)$. 
 
It is also possible to consider a third approach, namely a 
point by point comparison of 
$F_2$ both in the DIS and in the resonance
region. The latter can in fact be fitted to a smooth curve tracing 
the resonances
with a very high accuracy given by the increased precision of the new
Jefferson Lab measurements. They are then compared directly to 
DIS parameterisations $F_2$ at the same $x$ and $Q^2$ values.  
This procedure \cite{SIMO1} is much less sensitive to the elastic
contribution.  
All of the approaches described in this Section are necessary 
to interpret quantitatively the $Q^2$ dependence of duality and 
of its violation.

\subsection{Polarised structure function.}
The spin-dependent part of the polarised deep inelastic cross section is
given by: 
 
\begin{eqnarray}
\label{xsectpol}
& \frac{\displaystyle d^2\sigma}{\displaystyle dx dy}  = 
\displaystyle\frac{e^4}{2\pi^2 Q^2} (-H_e H_N)& \\ \nonumber
&  
\left[ \displaystyle \left( 1-\frac{y}{2}-\frac{y^2}{4}\gamma^2 \right)  
g_1(x,Q^2)  - \frac{y}{2} \gamma^2 g_2(x,Q^2) \right],
\end{eqnarray}
where $H_e$ and $H_N$ are the polarisations of the incident electron and
of the nucleon of the target, respectively.  
\indent
\par
Because of the mixing of $g_1$ and $g_2$, a precise determination of $g_1$
from a longitudinally polarised target alone is not possible.
The experimentally measured cross section asymmetries are the longitudinal
$A_{\parallel}$ and the transverse $A_{\perp}$ ones, formed from combining
data with opposite beam helicity:
\begin{equation}
A_\|=\frac{\sigma^{\downarrow\uparrow}-\sigma^{\uparrow\uparrow}}{\sigma^{\downarrow\uparrow}+\sigma^{\uparrow\uparrow}},
\hspace{1cm}
A_\perp=\frac{\sigma^{\downarrow\rightarrow}-\sigma^{\uparrow\rightarrow}}{\sigma^{\downarrow\rightarrow}+\sigma^{\uparrow\rightarrow}}.
\end{equation}
\indent
\par
The polarised structure functions are determined from these asymmetries:
\begin{eqnarray}
\label{g1g2}
g_1(x,Q^2)=\frac{F_1(x,Q^2)}{d'}\bigg[A_\|+\tan\theta/2 \cdot A_{\perp}\bigg], \nonumber \\
g_2(x,Q^2)=\frac{yF_1(x,Q^2)}{2d'}\bigg[\frac{E+E'\cos\theta}{E'\sin\theta}A_{\perp} -A_\| \bigg]
\end{eqnarray}
where $E'$ is the scattered lepton energy, 
$\theta$ is the scattering angle, 
$d'=[(1-\epsilon)(2-y)]/[y(1+\epsilon R(x,Q^2)]$, with $\epsilon$ being the 
degree of transverse polarization of the virtual photon and defined as 
$\epsilon^{-1}=1+2(1+\gamma^{-2})\tan^2(\theta/2)$.
\indent
\par
The virtual photon-absorption asymmetries $A_1$ and $A_2$ are related to the
measured asymmetries by:
\begin{eqnarray}
\label{Ameas}
A_\| = D (A_1+\eta A_2) \nonumber \\
A_{\perp} = d (A_2-\zeta A_1),
\end{eqnarray}
where $D$ is the depolarization factor 
$D= y(2-y)(1+\gamma^2y/2)/[y^2(1+\gamma^2)(1-2m_e^2/Q^2)+
    2(1-y-\gamma^2y^2/4)(1+R)]$, and $d=D \sqrt{2\epsilon/(1+\epsilon)}$, 
$\eta=2\gamma(1-y)/(2-y)$ and $\zeta=\eta (1+\epsilon)/2\epsilon$ are
kinematic factors.
\indent
\par
From the measured asymmetries $A_\|$ and $A_{\perp}$, the virtual photon 
asymmetries can be related to the photon absorption cross section of the
nucleon for a given $x$ and $Q^2$:
\begin{eqnarray}
\label{A1A2}
A_1 = \frac{\sigma_{1/2}-\sigma_{3/2}}{\sigma_{1/2}+\sigma_{3/2}}=\frac{\sigma_{TT}}{\sigma_T}=\frac{g_1-\gamma^2 g_2}{F_1} \nonumber \\
A_2 = \frac{2\sigma_{LT}}{\sigma_{1/2}+\sigma_{3/2}}=\frac{\sigma_{LT}}{\sigma_T}=\frac{\gamma (g_1+g_2)}{F_1}.
\end{eqnarray}
Here $\sigma_{1/2}$ and $\sigma_{3/2}$ are the virtual photo-absorption cross
section when the projection of the angular momentum of the photon-nucleon 
system along the incident photon direction is 1/2 or 3/2 respectively, 
$\sigma_{LT}$ is the interference term between the transverse and longitudinal 
photon-nucleon amplitudes, respectively given by 
$\sigma_T=(\sigma_{1/2}+\sigma_{3/2})/2$ and 
$\sigma_{TT}=(\sigma_{1/2}-\sigma_{3/2})/2$.
If only the longitudinal asymmetry is measured, it is necessary to make an
assumption for the asymmetry $A_2$.
\indent
\par
From the measured asymmetry it is possible to 
evaluate the polarised structure function $g_1$; 
in fact neglecting the term $\gamma^2 g_2$, Eq.(\ref{A1A2}) reduces to:
\begin{equation}
g_1(x,Q^2) \approx  A_1(x) F_1(x,Q^2).
\label{ourdef}
\end{equation}
This approximation is adequate for describing the data within their given 
accuracy. 

As for the unpolarised case, the twist expansion reads:
\begin{eqnarray}
\label{t-exp-pol} 
g_1(x,Q^2) = g_1^{LT}(x,Q^2) +
\frac{\widetilde{H}(x,Q^2)}{Q^2} + {\cal O}\left(1/Q^4 \right),
\end{eqnarray}
where, using Eqs.(\ref{R},\ref{TMC},\ref{ourdef}):
\begin{eqnarray}
\label{LT-pol} 
g_1^{LT}(x,Q^2) & = & F_2^{LT}(x,Q^2) A_1^{exp}(x) \nonumber \\ 
& &  \times (1+\gamma^2)/(2x(1+R^{exp})) ,  
\end{eqnarray}
where by the superscript ``exp'' we emphasize that we have used 
the experimental values for the quantities under consideration. 
   
We study parton-hadron 
duality by defining the integrals:
\begin{equation}
\widetilde{\Gamma}_1^{\mathrm{res}} = \int_{x_{\mathrm{min}}}^{x_{\mathrm{max}}}g_1^{\mathrm{res}}(x,Q^2)dx, 
\end{equation}
where $g_1^{\mathrm{res}}$ is obtained from the data in the resonance region, 
and 
\begin{subequations}
\begin{eqnarray}
\widetilde{\Gamma}_1^{LT} = \int_{x_{\mathrm{min}}}^{x_{\mathrm{max}}}g_1^{LT}(x,Q^2)dx
\\
\widetilde{\Gamma}_1^{HT} = \int_{x_{\mathrm{min}}}^{x_{\mathrm{max}}}
\bigg(g_1^{LT}(x,Q^2)+ \frac{\widetilde{H}(x,Q^2)}{Q^2}\bigg)dx
\end{eqnarray}
\end{subequations}
The ratios are given by: 
\begin{equation}
R_{\mathrm{pol}}^{LT}=\displaystyle\frac{\widetilde{\Gamma}_1^{\mathrm{res}}}{\widetilde{\Gamma}_1^{LT}} 
\hspace{0.5cm}
,
\hspace{0.5cm}
R_{\mathrm{pol}}^{HT}=\displaystyle\frac{\widetilde{\Gamma}_1^{\mathrm{res}}}{\widetilde{\Gamma}_1^{HT}}.
\label{Rdual_pol}
\end{equation}
As for the unpolarised case, duality is verified if either ratio is unity.\\

\section{Analysis and Interpretation of Data}


In this Section we present a quantitative analysis of the $Q^2$ dependence
of parton-hadron duality in both polarised and unpolarised $ep$ 
scattering. 
We take into account all current 
data in the resonance region, $1 \leq W^2 \leq 4$ GeV$^2$.
For the unpolarised case we used data obtained at Jefferson Lab in the
range $0.3 \leq Q^2 \leq 5$ GeV$^2$ \cite{CEBAF}, and data from SLAC 
(\cite{whit} and references therein) for $Q^2 \geq 4$ GeV$^2$.
For the polarised case there are only few experimental data in the 
resonance region. 
One set is part of the E143 data \cite{E143}, and it corresponds to 
$Q^2 =0.5$ and $1.2$ GeV$^2$. 
Another set is the one from HERMES \cite{HERMESPRL,Ale} in the range
$1.2 \leq Q^2 \leq 12$ GeV$^2$. 

In the polarised case the $Q^2$ dependence originates from 
the structure function $F_1$ and from the ratio $R$..
In the evaluation of the denominators of Eqs.(\ref{Rdual_pol}),
we used the SLAC global analysis~\cite{r1990} parameterisation for $R$. 
and, for $A_1$, a power law fit to the world DIS data at $x>$0.3: 
$A_1=x^{0.7}$, as already shown in Ref.~\cite{HERMESPRL}.
This parameterisation of $A_1$ is constrained to 1 at $x$=1 and it does not
depend on $Q^2$, as indicated by experimental data in this range~\cite{E155}. 

\subsection{Comparison with pQCD}
\begin{figure}[h]
   \begin{center}
      \mbox{
           \epsfysize=8cm
           \epsfbox{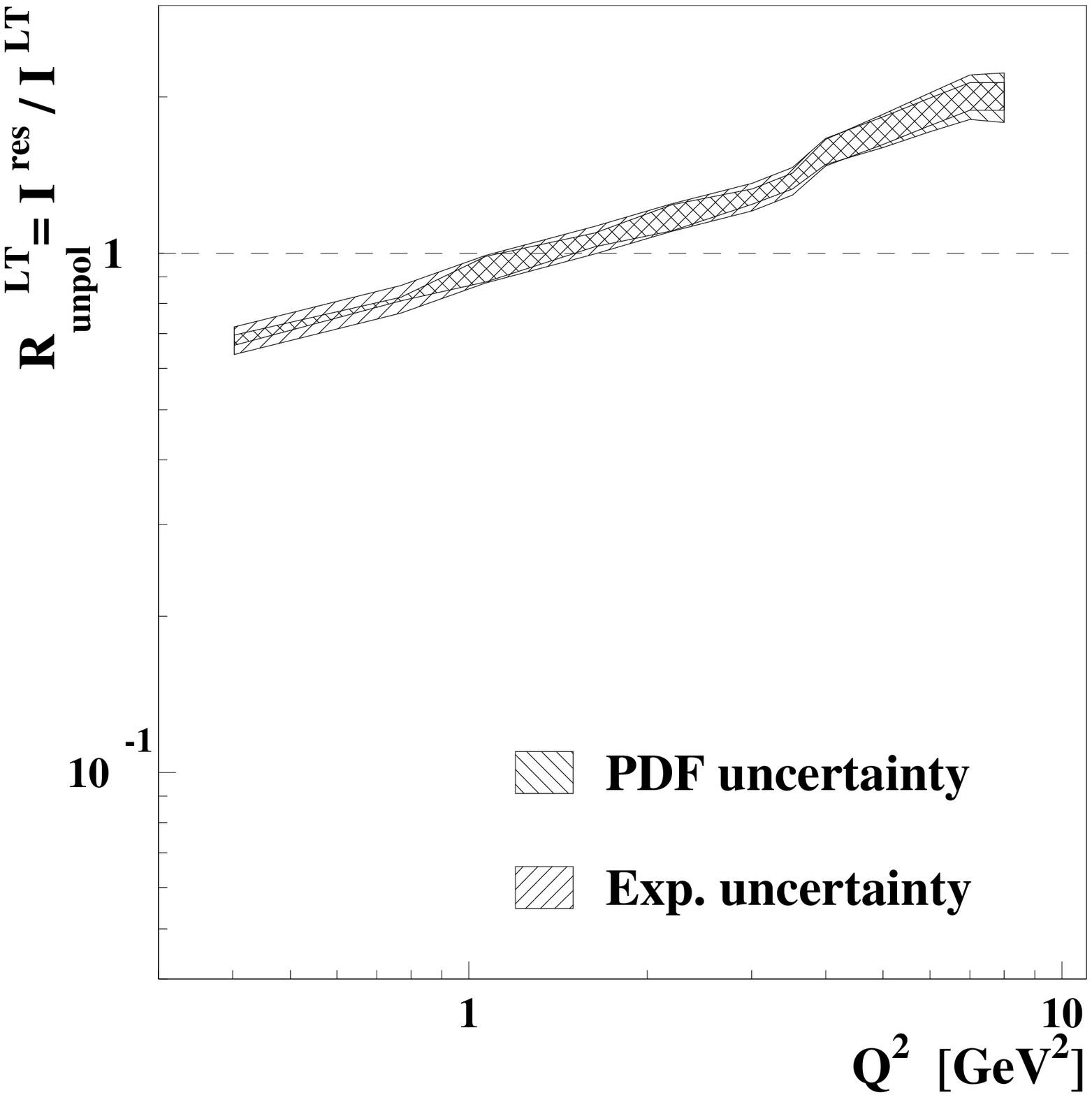}
           }
      \mbox{
           \epsfysize=8cm
	   \epsfbox{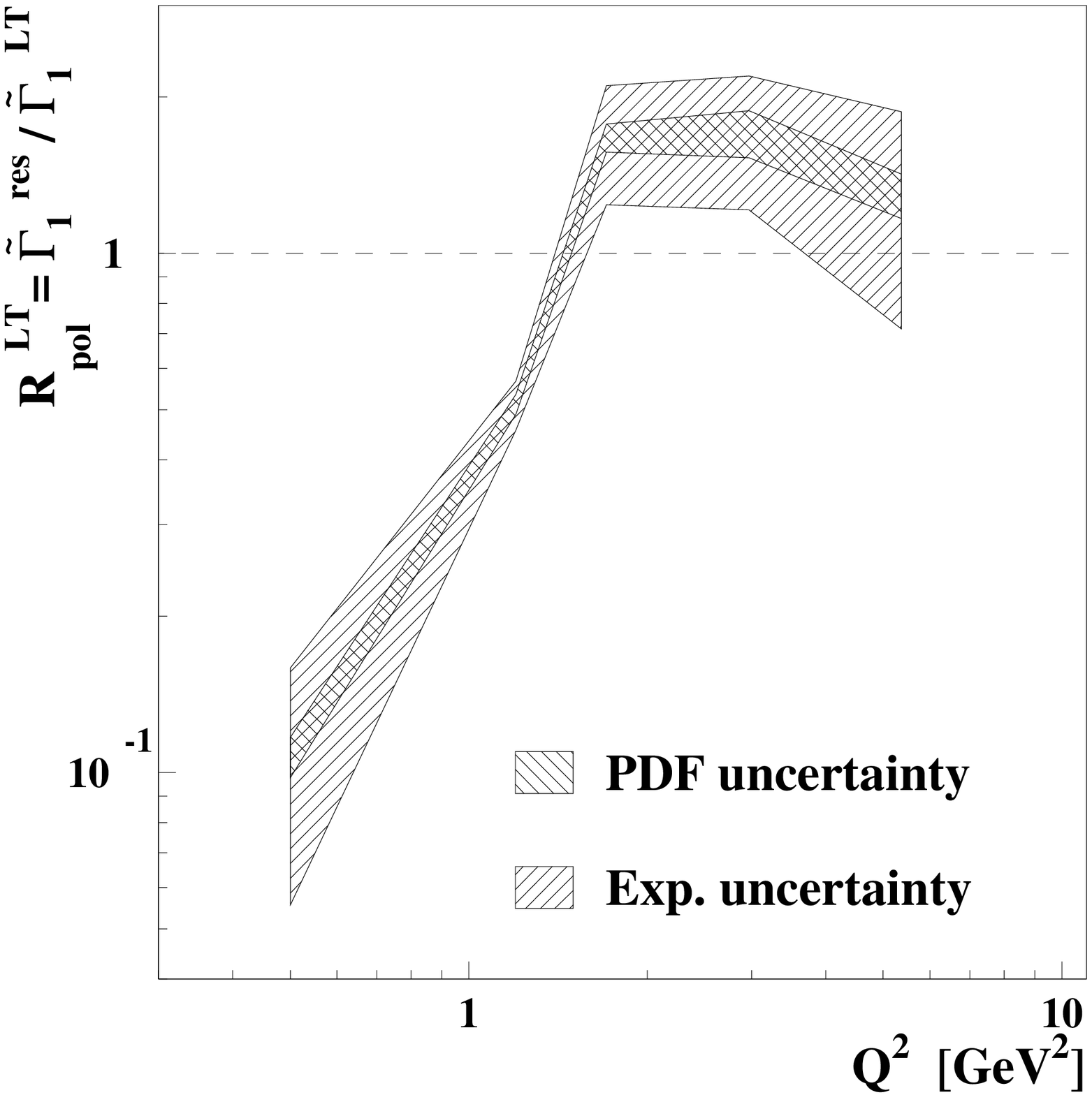}
           }
   \end{center}
\caption{Ratio between the integral of the structure function as measured in 
         the resonance region and as parameterised in the DIS region as a 
         function of $Q^2$. 
         The top panel refer to the unpolarised case, while
         the bottom panel to the polarised one.
         One band represents the theoretical uncertainty due to the use of 
         different LO and NLO parameterisations: MRST \cite{MRST},
         CTEQ \cite{CTEQ}, GRV94 \cite{GRV94}, GRV98 \cite{GRV98}. 
         The other band represents the experimental uncertainty, that
         is the sum in quadrature of the statistical and systematic 
         uncertainties of the data in the resonance region.    	 
}
\label{PDF}
\end{figure}

The unpolarised structure function $F_2$ was first 
evaluated from dynamical parameterisations, coming from the Parton 
Distribution Functions (PDFs):
MRST~\cite{MRST}, CTEQ~\cite{CTEQ}, GRV94~\cite{GRV94} and GRV98~\cite{GRV98}.
The last two parameterisations have been evaluated at Leading Order (LO) and
at Next to Leading Order (NLO).
All of them are pure DIS parameterisations and they were extended to the 
measured $x$ and $Q^2$ ranges by pQCD evolution.
The $Q^2$ evolution of the polarised parton densities is governed by the 
Dokshitzer-Gribov-Lipatov-Altarelli-Parisi (DGLAP)~\cite{DGLAP} equations.
The results are shown in Fig.\ref{PDF}, in the top panel for the
unpolarised data and in the bottom panel for the polarised data.
The uncertainty due to the use of different LO and NLO
parameterisations is represented with a band; the other band represents 
the experimental uncertainty, calculated as the sum in quadrature of
the statistical and systematic uncertainties of the data in the resonance 
region. 

Because the structure functions are dominated by large $x$ kinematics,
DGLAP evolution proceeds only through Non-Singlet (NS) distributions. 
This explains why there is very little uncertainty in the extrapolation of 
the initial pQCD distribution to the low values of 
$W^2$ considered and the small difference between LO and NLO evolution.

Parton-hadron duality is not fulfilled by using solely the PDFs up to
NLO in both the unpolarised and polarised structure functions
$F_2$ and $g_1$.
However it is possible to see a different behavior between
$R_{\mathrm{unpol}}$ and $R_{\mathrm{pol}}$.
In the unpolarised case the ratio is increasing with $Q^2$, but for
the polarised case the situation is different: while at low $Q^2$ 
the ratio is significantly below unity and shows a strong increase 
with $Q^2$, at higher $Q^2$ the ratio derived from HERMES is above unity and
they appear to be independent of $Q^2$ within error bars.

\begin{figure}[h]
   \begin{center}
      \mbox{
           \epsfysize=8cm
           \epsfbox{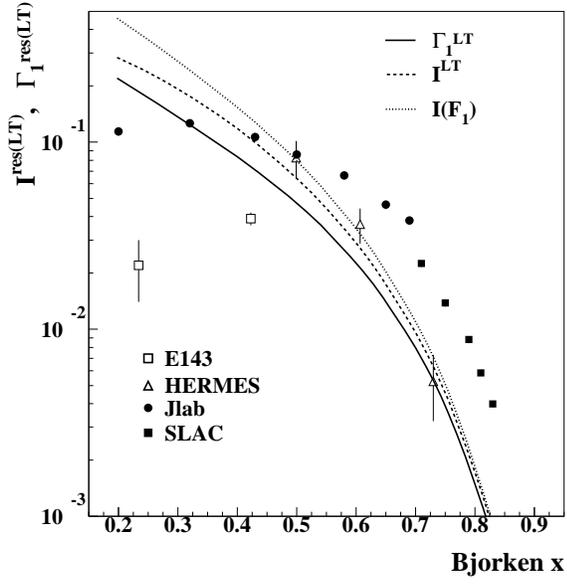}
           }
   \end{center}
\caption{The integrals $I^{res}$, Eq.(\protect\ref{Iexp}), and
$\widetilde{\Gamma}_1^{res}$, Eq.(\protect\ref{Iexp}) 
plotted vs. the average value of Bjorken x defined in the
text, obtained using the data from
Refs.~\cite{E143,HERMESPRL}. Experimental data are compared to the 
integrals of the valence component of the  
structure functions $g_1$ (full line), $F_2$ (dashed line),   
and $F_1$ (dotted line), calculated using NLO parameterisations.
}
\label{pol_moment}
\end{figure}

In Figure \ref{pol_moment} we further illustrate the origin of this behavior
by plotting separately the numerator (data points) and the denominator
(``theoretical'' curves representing a pQCD based parameterisation), 
of the ratios $R_{unpol}$ and $R_{pol}$, 
respectively. We also plot the integral of 
$F_1$ (dotted line) in order to show the effect of both $A_1$ and of 
the $Q^2$-dependent
factors that come into play in the definition of $g_1$. 
All quantities are plotted vs. $x \equiv \langle x \rangle$, 
{\it i.e.} the average value of Bjorken $x$ for each spectrum, 
defined in Section II. 
Notice that the value of $\langle x \rangle$ increases with
$Q^2$. The trends in the figure suggest therefore that similarly to
what observed in DIS, in the resonance 
region there are corrections beyond DGLAP evolution that are positive at
large $x$, and negative at smaller $x$, the threshold being 
defined by: $x \approx 0.33 - 0.43 $ and 
$Q^2 \approx 1 $ GeV$^2$.
However, while these corrections are comparable in size for both 
the polarised and unpolarised case at large $x$, at low $x$ there seems
to be a much larger non-perturbative effect for the polariseddata. 
While possible explanations have been suggested {\it e.g.} in \cite{CloMel}, 
it is clear that more data in this region would help disentangling
the $Q^2$ dependence and the possible size of the non-perturbative 
effects.        

\subsection{Comparison with Phenomenological Parameterisations}
The unpolarised structure function $F_2$ has been 
evaluated from three different phenomenological fits to DIS data 
\cite{nmcp8,ALLM,BOYA} and scaled to the same $Q^2$ values as for 
$I^{\mathrm{res}}$ and $\widetilde{\Gamma}_1^{\mathrm{res}}$ to take into 
account the large effect of scaling violation at large $x$.

The parameterisation from Ref.~\cite{BOYA} is a modification
of the PDF ~\cite{GRV94} that purports to include target mass effects through
an effective change of variable, higher twist effects at high $x$, and a 
factor that enables an extension of the fit down to the
photoproduction limit.
The ALLM parameterisation \cite{ALLM}
is based on a reggeon and pomeron exchanges and it
was constructed for the high $W^2$ limit. 
We included it because it can be extended to
very low $Q^2$ values.  
The NMC parameterisation \cite{nmcp8} includes a fit of $HT$ terms using world 
data with $W^2>10$ GeV$^2$.

\begin{figure}[h]
   \begin{center}
      \mbox{
           \epsfysize=8cm
           \epsfbox{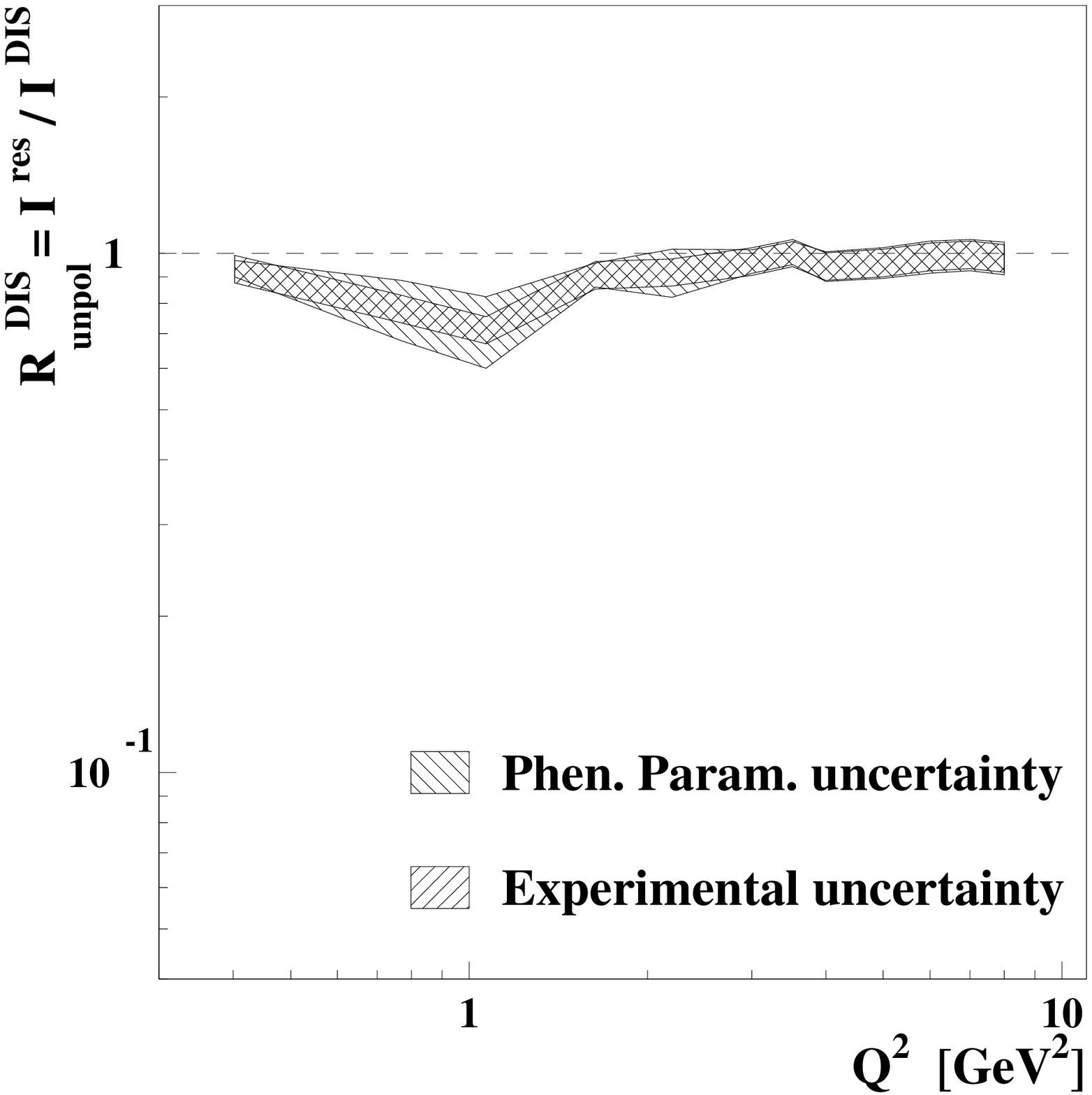}
           }
      \mbox{
           \epsfysize=8cm
           \epsfbox{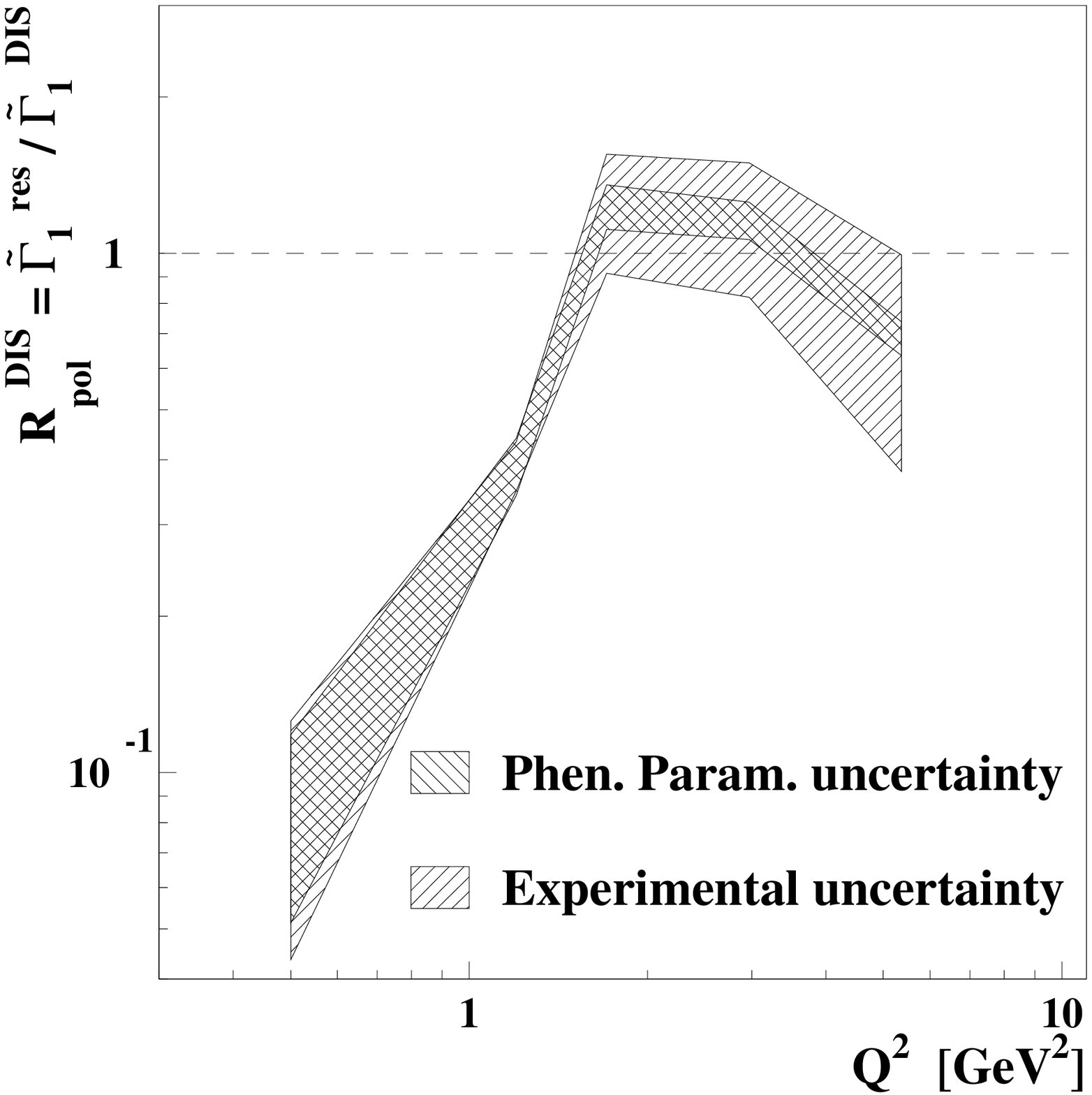}
           }
   \end{center}
\caption{Ratio between the integral of the structure function measured in 
         the resonance region and as parameterised in the DIS region 
         \cite{BOYA,nmcp8,ALLM} as a function of $Q^2$.
         The top panel refers to the unpolarised case, while
         the bottom panel to the polarised one.
         Notations are as in Fig.\ref{PDF}.}
\label{phen}
\end{figure}

The ratios 
$R_{\mathrm{unpol}}^{\mathrm{DIS}} = I^{\mathrm{res}}/I^{\mathrm{DIS}}$ and
$R_{\mathrm{pol}}^{\mathrm{DIS}} = \widetilde{\Gamma}_1^{\mathrm{res}}/\widetilde{\Gamma}_1^{\mathrm{DIS}}$, where $I^{\mathrm{DIS}}$ and $\widetilde{\Gamma}_1^{\mathrm{DIS}}$ are the integrals calculated with these phenomenological 
parameterisations, are shown in Fig.\ref{phen} in the top and bottom panel,
respectively, for several $Q^2$-values.

In the unpolarised case the slope of the ratio is less evident compared to 
the previous method shown in Fig.\ref{PDF}
and the ratio is well consistent with unity for $Q^2 >$ 2
GeV$^2$.
In the polarised case, the ratio at higher $Q^2$ derived from HERMES data is 
consistent with unity inside the experimental errors and still independent 
of $Q^2$.
However, at low $Q^2$ the uncertainty on the parameterisation is bigger
than what found from the PDFs.

Since these phenomenological parameterisations are obtained by fitting deep-inelastic data
even in the low $Q^2$ region, they can implicitely include 
non-perturbative effects and this may explain the ``observation of duality''.

\subsection{Size of Non-perturbative Contributions}
In order to understand the nature of the remaining $Q^2$ dependence that cannot
be described by NLO pQCD evolution, we studied the effect of TMC and 
LxR on the ratios $R_{unpol}^{HT}$ and $R_{pol}^{HT}$. 
The analysis was performed by using $x$ as an integration variable, which 
avoids the ambiguities associated to the usage of 
other {\it ad hoc} kinematical variables.
We used standard input parametrizations with initial scale $Q_o^2 = 1 $ GeV$^2$. 
Once TMC and LxR have been subtracted from the data, and assuming the validity 
of the twist expansion, 
Eqs.(\ref{t-exp},\ref{t-exp-pol}) in this region, one can 
interpret any remaining discrepancy in terms of HTs. 

We notice that although we did not consider NNLO calculations, these are not expected
to alter substantially our extraction since,
differently from what seen originally in the case of $F_3$, these have been proven 
to give a relatively small contribution to $F_2$ \cite{Ale1}.  

The value of $\alpha_S(M_Z^2)$ that was used in our calculations  
corresponds to the one given for the DIS 
parameterisations.
It has been long noticed that 
a correlation exists between $\alpha_S$ and the extracted values of the HTs
(see \cite{Penn} and references therein and the recent highly accurate 
determination in 
Ref.~\cite{sia_alphas}). It is exactly because of this correlation that 
we keep its value fixed from evaluations in a region where the HTs contribution is negligible.
This statement is equivalent to saying that $\alpha_S$ cannot be extracted reliably
from large $x$ data.

TMC have been evaluated using Eq.(\ref{TMC}) for the unpolarised case, and 
Eqs.(\ref{TMC},\ref{LT-pol}) for the polarised data.     
Although this procedure disregards parton off-shell effects that might be important
in the resonance region (see Refs.\cite{FraGun,ioanaPRD}), we emphasize here its power 
expansion character, and we set as a limiting condition for its validity, that the 
inequality: $x^2M^2/Q^2 < 1$, be verified \cite{SIMO1}. Therefore, current
treatments of TMC in the resonance region are uncertain for values of 
$Q^2 \lesssim 1.5$ GeV$^2$.
  
LxR effects arise formally from terms containing powers of 
$\ln (1-z)$, $z$ being the longitudinal 
variable in the evolution equations, that are present in 
the Wilson coefficient functions $C(z)$. 
The latter relate the parton distributions to {\it e.g.} 
the structure function $F_2$, according to:
\begin{equation}
F_2^{NS}(x,Q^2)  = \frac{\alpha_s}{2\pi} \sum_q \int_x^1 dz \, C_{NS}(z) \, q_{NS}(x/z,Q^2), 
\end{equation}   
where we have considered only the non-singlet (NS) contribution to $F_2$ since 
only valence quarks distributions are relevant in our kinematics. 
The logarithmic terms in $C_{NS}(z)$ become very large at large $x$, and they need to be 
resummed to all orders in $\alpha_S$. 
This can be accomplished by  
noticing that the correct kinematical variable that determines the 
phase space for the radiation of gluons
at large $x$, is $\widetilde{W}^2 = Q^2(1-z)/z$, 
instead of $Q^2$ \cite{BroLep,Ama}. 
As a result, the argument of the strong coupling constant becomes $z$-dependent: 
$\alpha_S(Q^2) \rightarrow \alpha_S(Q^2 (1-z)/z)$ (see \cite{Rob} and references therein). 
In this procedure, however, an ambiguity is introduced, related to the need of continuing 
the value of $\alpha_S$  
for low values of its argument, {\it i.e.} for $z$ very close to $1$ \cite{PenRos}. 
The size of this ambiguity could be of the same order of the HT corrections. 
Nevertheless, our evaluation is largely free from this problem because of the particular kinematical 
conditions in the resonance region. We are in fact studying the structure functions 
at {\it fixed}  $W^2$, in between $1 \leq W^2 \leq 4$ GeV$^2$. Consequently $Q^2$  
increases with $x$. This softens the ambiguity in $\alpha_S$, and renders 
our procedure reliable for the extraction of HT terms. 
We illustrate this situation in Fig.6 where we plot 
the value of $\alpha_S$ at $Q^2=10$ GeV$^2$, and we compare it with the resummed value
in the resonance region, at $Q^2(1-z)/z$ for a fixed average $Q^2$, and at  $Q^2(x)(1-z)/z$,
with $Q^2(x) =W^2 \langle x \rangle /(1-\langle x \rangle)$, and $\langle x \rangle = 0.83$.

All of the effects described in this Section are summarized in
the upper panel of Fig.\ref{theor1}.  
In the figure we plot the ratio $R_{unpol}^{LT}$, 
from Eq.(\ref{RdualLT}), where the numerator is obtained from the 
experimental data, while the denominator includes the different components 
of our analysis, one by one. 
For unpolarised scattering we find that TMC and LxR diminish considerably
the space left for HT contributions. 
The contribution of TMC is large at the largest values of $Q^2$ because these correspond also to
large $x$ values. Moreover, the effect of TMC is larger than the one of LxR.
We have excluded from our analysis the lowest data point at $Q^2\approx 0.4$ GeV$^2$ 
because of the high uncertainty in both the pQCD calculation and the subtraction of 
TMC. Also, the pQCD calculations at $Q^2\approx 1$ GeV$^2$ differ from the ones obtained
by using the available set of parametrizations perhaps because the latter are 
extrapolated well beyond their limit of validity. 

Similarly, in polarised scattering the inclusion of TMC and LxR decreases 
the ratio $R_{pol}^{LT}$
(Fig.\ref{theor1}, bottom panel). 
However, in this case these effects are included almost completely 
within the error bars. 
We conclude that duality is strongly violated at $Q^2 < 1.7$ GeV$^2$.

The difference between unpolarised and polarised scattering at low $Q^2$ can be 
attributed {\it e.g.} to unmeasured, so far, $Q^2$ dependent effects, both 
in the asymmetry, $A_1$,  and in $g_2$. Furthermore, a full treatment
of the $Q^2$ dependence would require both a more accurate knowledge of the ratio 
$R$ in the resonance region, and a simultaneous evaluation of $g_2$.
The present mismatch between the 
unpolarised and polarised low $Q^2$ 
behavior might indicate that factorization is 
broken differently for the two processes, and that the universality
of partonic descriptions no longer holds.          

In Figures \ref{theorx}, \ref{HTcomp} we address explicitely the question of the size of the HT corrections. We define them for $F_2$ as:
\begin{subequations}
\begin{eqnarray}
\label{CHT}
H(x,Q^2) & = & Q^2 \left(F_2^{\mathrm{res}}(x,Q^2) - F_2^{\mathrm{LT}} \right) \\ 
C_{HT}(x) & = & \frac{H(x,Q^2)}{F_2^{pQCD}(x/Q^2)}  \nonumber \\  
& & \equiv Q^2 \frac{F_2^{\mathrm{res}}(x,Q^2) - F_2^{LT}}{F_2^{\mathrm{LT}}}.   
\end{eqnarray}
\end{subequations}
A similar expression is assumed for $g_1$. 
$C_{HT}$ is the so-called factorised form obtained by assuming that
the $Q^2$ dependences of the LT and of the HT parts are similar and therefore
they cancel out in the ratio. Although the anomalous 
dimensions of the HT part could in principle be different, such a discrepancy 
has not been found so far in accurate analyses of DIS data.   
The HT coefficient, $C_{HT}$ has been evaluated for the three cases listed also in Fig.\ref{theor1},
namely with respect to the NLO pQCD calculation, to NLO+TMC 
and to NLO+TMC+LxR. 
The values of $1 + C_{HT}/Q^2$ are plotted in Fig.\ref{theorx} (upper panel) 
as a function of the average value of $x$ 
for each spectrum. One can see that the NLO+TMC+LxR analysis yields very small values for $C_{HT}$ 
in the whole range of $x$. 
Furthermore, the extracted values are consistent 
with the ones obtained in Ref.\cite{SIMO1} using a different method, however the 
present extraction method gives more accurate results. 
Because of the increased precision of our analysis, we are able to disentangle 
the different effects from both TMC and LxR. 
   
In the polarised case (Fig.8, lower panel) the HTs are small within the given precision, 
for $Q^2 > 1.7$ GeV$^2$, 
but they appear to drop dramatically below zero for lower $Q^2$ values. 
The inclusion of TMC and LxR
renders these terms consistent with zero at the larger $Q^2$ values, but it does not 
modify substantially their behavior at lower $Q^2$. 

In Fig.\ref{HTcomp} (upper panel) 
we compare our results in the unpolarised case to other current extractions of the same 
quantity. These are: {\it i)} the extractions from DIS data, 
performed with the cut: $W^2 > 10 $ GeV$^2$ \cite{VM,MRST_HT,Botje}; 
{\it ii)} the recent DIS evaluation by S. Alekhin \cite{Ale1} using 
a cut on $W^2 > 4$ GeV$^2$, and including both TMC and NNLO; {\it iii)} 
the results obtained within a fixed $W^2$ framework in Ref.~\cite{SIMO1}, 
including both TMC and LxR. 
We notice that results obtained in Ref.~\cite{ScScSt} in the deep inelastic region 
also including both TMC and LxR 
yield small HT coefficients, consistent with the ones 
found in Ref.~\cite{SIMO1}.
However, while most of the suppression of the HT in the resonance 
region is attributed to TMC, 
in \cite{ScScSt} the contribution of TMC is small and the 
suppression is dominated by LxR. In other words, the $Q^2$ behavior 
in the DIS and resonance 
regions seems to be dominated by different effects.   
In Fig.9 (lower panel) we compare the HT coefficients in the unpolarised and polarised
case. One can notice a considerable discrepancy at $Q^2 \leq 1.7$ GeV$^2$. 

Our detailed extraction of both the $Q^2$ dependence and the HTs in the resonance region 
establishes a background for understanding the transition between partonic 
and hadronic degrees of freedom. 
In particular, we seem to be detecting a region 
where the twist expansion breaks down, 
and at the same time, the data seem to be still far from 
the $Q^2 \rightarrow 0$ limit, where theoretical predictions can be made 
\cite{JiO}.    
This breakpoint is marked, for instance, by the discrepancy between 
polarised and unpolarised scattering at $Q^2 \lesssim 1.7$ GeV$^2$. 
More studies addressing this region will be pursued in the future, 
some of which are also mentioned in \cite{SIMO1,SIMO_elba}. 
In particular, a breakdown of the twist expansion 
can be interpreted in terms of the dominance of multi-parton 
configurations over single parton contributions 
in the scattering process. 
In order to confirm this picture it will be necessary to both extend the
studies of the twist expansion, including the possible $Q^2$ dependence of 
the HT coefficients and terms of order ${\cal O}(1/Q^4)$, and
to perform duality studies in semi-inclusive experiments. 
Finally, a number of scenarios have been studied in \cite{CloMel}, 
that consider $SU(6)$ quark parton model breaking effects 
and that preserve duality in polarised scattering. 
The results of Ref.\cite{CloMel} are consistent only with the larger $Q^2$ 
behavior of the data, although they do not address explicitely the question 
of duality violations. 

\section{Conclusions}
In summary, we presented a study of parton-hadron duality in both unpolarised 
and polarised scattering. 
The latter was obtained by using the first experimental data for the
polarised structure function of the proton $g_1^{\mathrm p}(x)$, 
for $Q^2$ values larger than 1.7 GeV$^2$. 
Parton-hadron duality was analysed within a QCD context. A pQCD NLO analysis
including target mass corrections and large $x$ resummation effects
was extended to the integrals of both unpolarised and polarised structure
functions in the resonance region. 
Within our context, duality is satisfied if the pQCD calculations 
agree with the data, modulo higher twist contributions consistent with the
twist expansion. 
Although the latter are found to be very small for the unpolarised 
structure function, we do not conclude that parton-hadron duality
holds straightforwardly. 
On the contrary, our findings seem to unveil 
a richer $Q^2$ dynamics both at $x \rightarrow 1$ and at small $x$. 
This observation is substantiated by the fact that duality 
holds when comparing data in the resonance region with 
phenomenological fits which contain some additional 
``non conventional'' $Q^2$ dependence, beyond what predicted 
by the twist expansion. 
Most importantly, the coefficient of the HTs extracted using 
data in the resonance region only, is smaller, 
and therefore not consistent with the one extracted in the DIS region.
Finally, while the size of the HT contributions is comparable in both
polarised and unpolarised scattering at larger 
$x$ and $Q^2$ values, at low $x$ and $Q^2$
we find large negative higher twists only in the polarised case.   

\begin{acknowledgments}
We are indebted 
to S. Alekhin for discussions and for comunications on his
calculations prior to publication. We thank W. Melnitchouk for 
discussions. S.L. thanks the Gruppo III of I.N.F.N.
at the Laboratori Nazionali di Frascati where this paper was 
completed, for their warm hospitality,
and for the many lively discussions.
This work was completed under the U.S. 
Department of Energy grant no. DE-FG02-01ER41200.
\end{acknowledgments}



\newpage

\begin{figure}[h]
\label{alpha_s}
   \begin{center}
      \mbox{
           \epsfysize=8cm
           \epsfbox{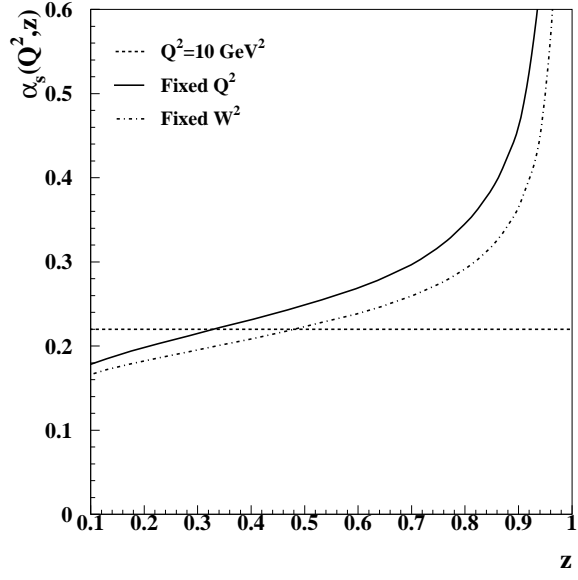}
           }
   \end{center}
\caption{$\alpha_S$ calculated to NLO for different forms of its argument: 
at $Q^2 = 10$ GeV$^2$ (dashes),
at $Q^2 \rightarrow Q^2(1-z)/z$ (full line), and $Q^2 \rightarrow W^2 x/(1-x) (1-z)/z$ 
(dot-dashed line). Calculations including LxR in the resonance region use the latter form.}
\end{figure}
\begin{figure}[h]
   \begin{center}
      \mbox{
           \epsfysize=8cm
           \epsfbox{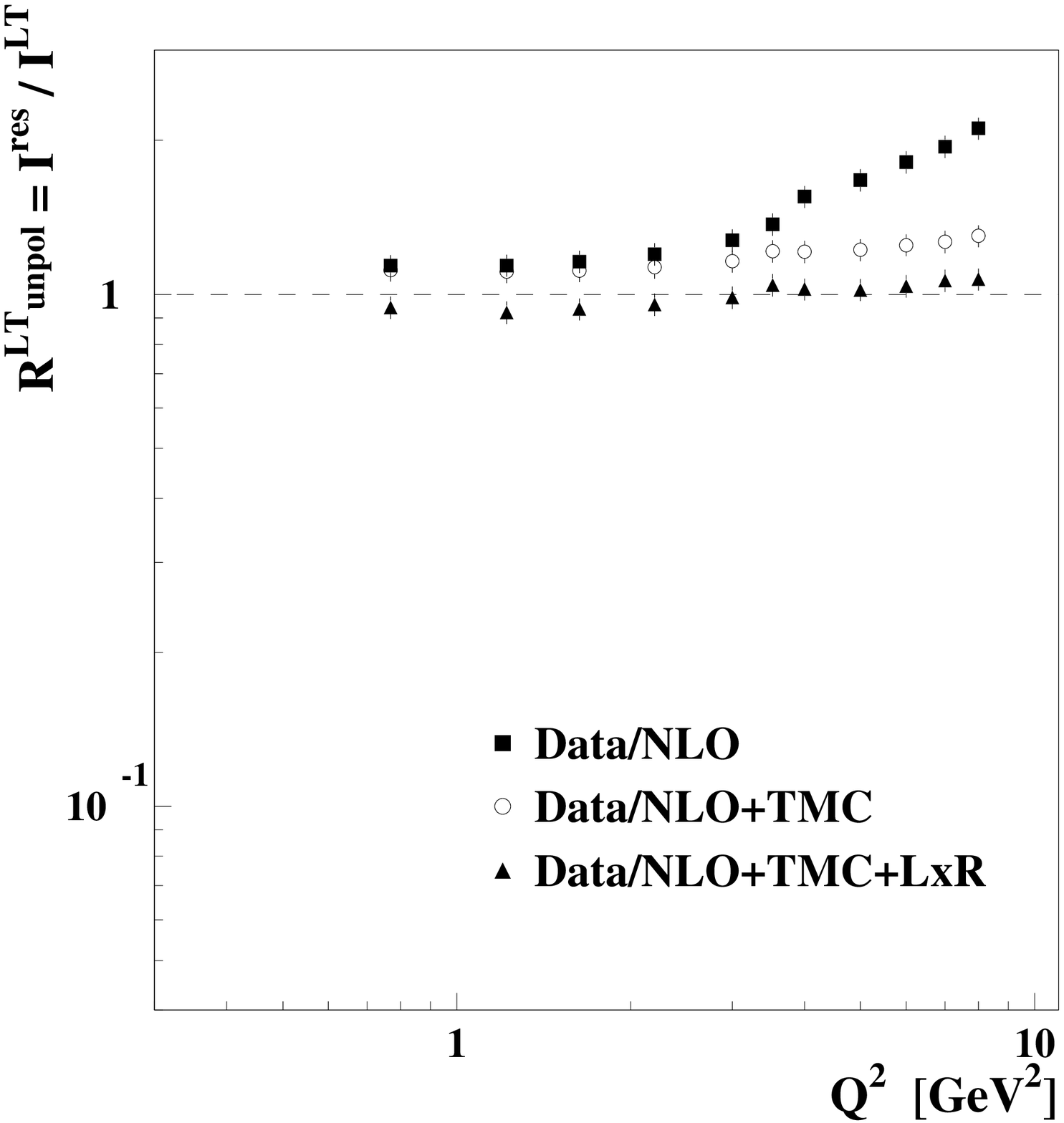}
           }
      \mbox{
           \epsfysize=8cm
           \epsfbox{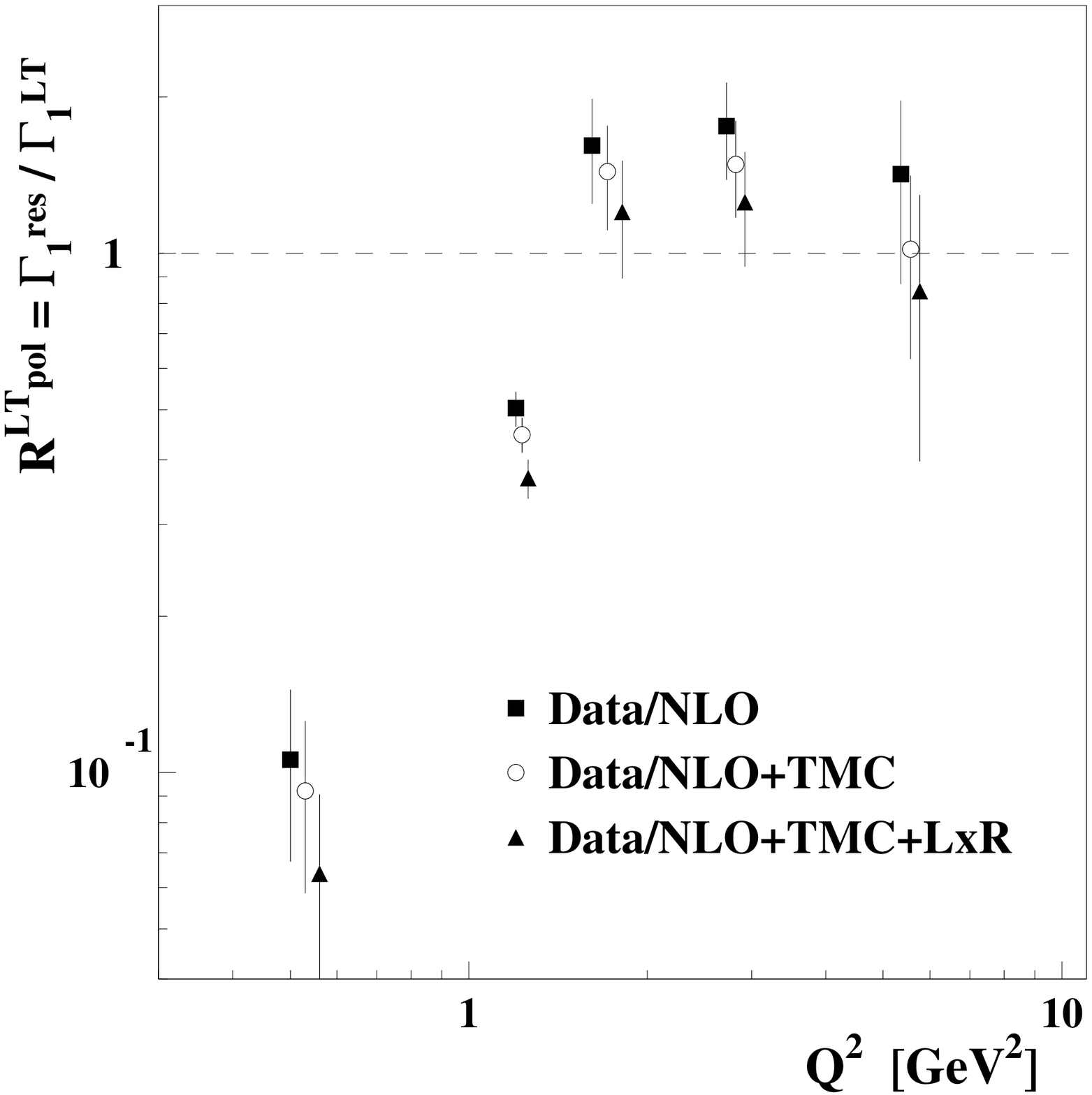}
           }
   \end{center}
\caption{Ratio between the integrals of the measured structure functions 
         and the calculated ones plotted as a  
         function of $Q^2$.  
         The calculation includes one by one the effects of NLO pQCD
         (squares), 
         TMC (open circles) and LxR (triangles),  
         The top panel refers to the unpolarised case, while
         the bottom panel to the polarised one.}
\label{theor1}
\end{figure}

\begin{figure}[h]
   \begin{center}
      \mbox{
           \epsfysize=8cm
           \epsfbox{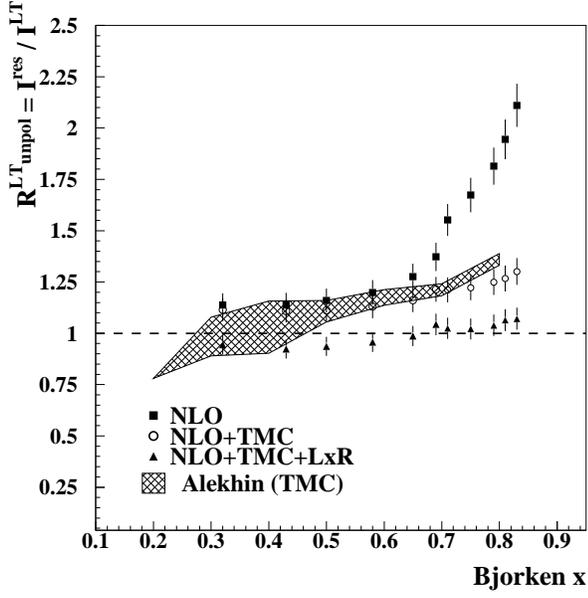}
           }
      \mbox{
           \epsfysize=8cm
           \epsfbox{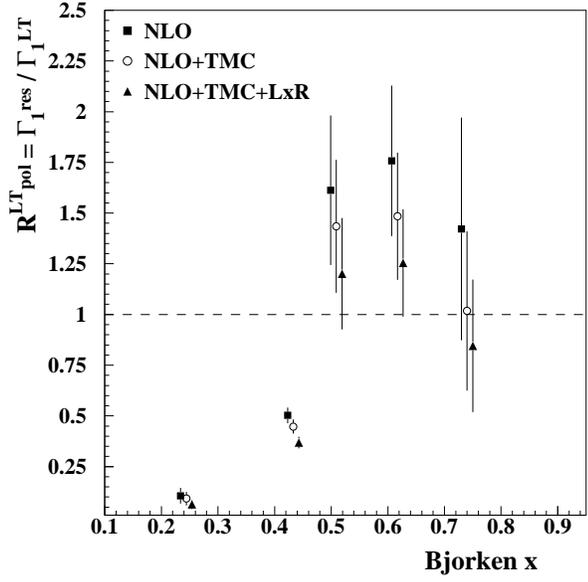}
           }
   \end{center}
\caption{HT coefficients extracted in the resonance region according 
         to Eq.(\protect\ref{CHT}). Shown in the figure is the quantity:
         $1+C_{HT}(x)/Q^2$.
         The top panel refers to the unpolarised case, where we show 
         the HT term obtained by considering only the NLO calculation (squares); the 
         effect subtracting TMC (open circles); and the effects of subtracting both 
         TMC and LxR (triangles). We show for comparison the values obtained from 
         the coefficient $H$ obtained in Ref.\protect\cite{Ale1} using DIS data and 
         including the effect of TMC.  
         The bottom panel refers to the HT coefficient in the polarised case. Notations
         are the same as in the upper panel.}
         
\label{theorx}
\end{figure}
\begin{figure}[h]
   \begin{center}
      \mbox{
           \epsfysize=8cm
           \epsfbox{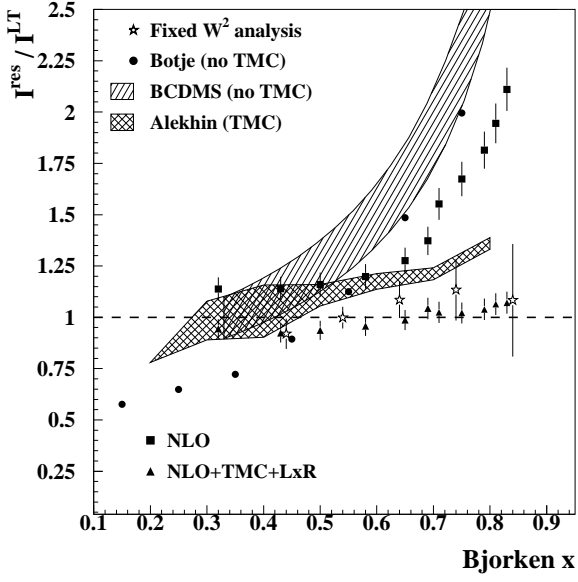}
           }
      \mbox{
           \epsfysize=8cm
           \epsfbox{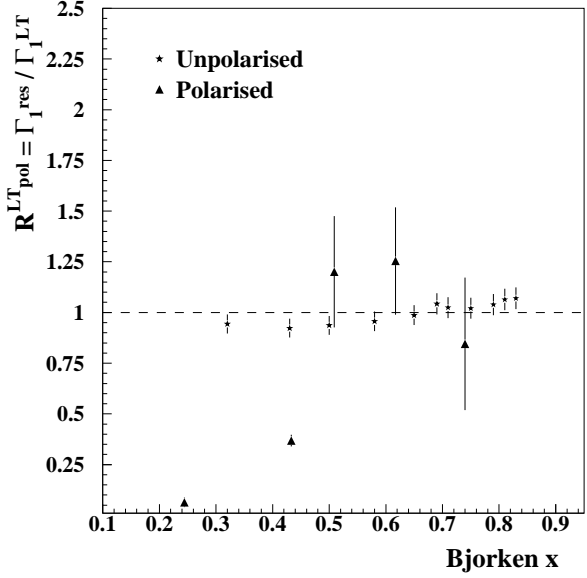}
           }
   \end{center}
\caption{Comparison of the HT coefficient displayed in Fig.8, with other 
         extractions (upper panel). The triangles and squares are the same as in Fig.8 and they 
         represent our determination in the resonance region. Our results are compared with 
         extractions using DIS data only. The striped hatched area corresponds to the 
         early extraction of Ref.\cite{VM}. The full dots are the central values of the 
         extractions in Refs.\cite{MRST_HT} and \cite{Botje}. These are compared with the 
         more recent extraction of Ref.\cite{Ale1} which includes also TMC. 
         Results obtained in the resonance region, in the fixed $W^2$ analysis 
         of Ref.\cite{SIMO1} are also shown (stars). In the bottom panel we show the comparison
         between the HTs in the resonance region, in the polarised (triangles) 
         and in the unpolarised (stars) cases.}
\label{HTcomp}
\end{figure}

\end{document}